\documentstyle{article}

\newcommand{\Section}[1]{\section{#1}\setcounter{equation}{0}}

\def\QED{\hbox{\kern 1pt\vrule width 3pt height 7pt}}
\def\p{\partial}
\def\db{\bar\partial}

\def\d{\delta}
\def\dg{\dagger}

\def\b{\bar}

\def\be{\begin{equation}}
\def\bea{\begin{eqnarray}}

\def\nn{\nonumber}

\def\l{\label}
\def\ee{\end{equation}}
\def\eea{\end{eqnarray}}
\def\C{\rm {I\kern-.520em C}}

\def\o{\over}
\begin{document}
\begin{titlepage}
\hfill
\vbox{
    \halign{#\hfil         \cr
           hep-th/0008120 \cr
           IPM/P2000/026   \cr
           } 
      }  
\vspace*{3mm}
\begin{center}
{\large \bf Gauged Noncommutative Wess-Zumino-Witten Models }
\vskip .5in
{\bf Amir Masoud Ghezelbash$^{*,\dagger,}$}\footnote{
amasoud@theory.ipm.ac.ir},
{\bf Shahrokh Parvizi$^{\dagger,}$}\footnote{
parvizi@theory.ipm.ac.ir}\\

\vskip .25in
{\em
$^*$Department of Physics, Az-zahra University,
Tehran 19834, Iran.}\\
{\em
$^\dagger$Institute for Studies in Theoretical Physics and Mathematics, \\
P.O. Box 19395-5531, Tehran, Iran.}\\
\vskip .5in
\end{center}
\begin{abstract}
We investigate the Kac-Moody algebra of noncommutative
Wess-Zumino-Witten model and find its structure to be the same as the
commutative case. Various kinds of gauged
noncommutative WZW models are constructed. In particular, noncommutative
$U(2)/U(1)$ WZW model is studied and by integrating
out the gauge fields, we obtain a noncommutative non-linear $\sigma$-model.
\end{abstract}
\end{titlepage}
\newpage
\def \dbar {\bar \partial}
\def \d  {\partial}
\Section{Introduction}
Noncommutative field theory has emerged from
string theory
in certain backgrounds \cite{CONNES,CHEUNG,SEIBERG,ARDALAN}. The
noncommutativity of space is defined by
the relation,
\be [x^\mu ,x^\nu ]=i\theta ^{\mu\nu}
\ee
where $\theta ^{\mu\nu}$ is a second rank antisymmetric real constant tensor.
The function algebra in the noncommutative space is defined by the
noncommutative and associative Moyal $\star $-product,
\be
(f\star g)(x)=e^{{i\over 2}\theta _{\mu\nu}{\partial\over {\partial \xi ^\mu}}
{\partial\over {\partial \eta ^\nu}}}f(\xi)g(\eta)|_{\xi =\eta =x}.
\ee

A noncommutative field theory is simply obtained by replacing
ordinary multiplication of functions by the Moyal $\star $-product.
An interesting field theory whose noncommutative version would be of
interest is the WZW model.
In \cite{DABROWSKI}, a noncommutative
non-linear $\sigma$-model has been studied and an infinite dimensional
symmetry
is found. They also derived some properties of noncommutative
WZW model. In \cite{FURUTA}, the $\beta$-function of the $U(N)$ noncommutative
WZW model was calculated and found to be the same as that of ordinary
commutative WZW model. Hence, the conformal symmetry in certain fixed points
is recovered. In \cite{MORENO} and \cite{CHU}, the
derivation of noncommutative WZW action from a gauge theory was carried out.
The connection between noncommutative two-dimensional fermion models and
noncommutative WZW models was studied in \cite{NEW,NEWW}.

In this letter, we study the noncommutative two-dimensional field theory
of WZW model and its gauged versions.  In section 2,
we calculate the current Kac-Moody algebra of the noncommutative WZW model
and find that the structure and central charge of the algebra are the same
as commutative WZW model.  In
section 3, we construct different versions of gauged noncommutative WZW models.
As an example, we consider the axial gauged $U(2)$ WZW model by its diagonal
$U(1)$ subgroup. The obtained gauged action contains infinite derivatives
in its $\star$ structure and hence is a nonlocal field theory. Integration
over the gauge fields requires solving an integral equation which we solve
by perturbative expansion in $\theta$. The result is a
noncommutative non-linear $\sigma$-model, which may contains singular
structures or a black hole.

\Section{The Current Algebra of Noncommutative WZW Model}
The action of the noncommutative WZW model is \cite{MORENO}:
\be \l{action1}
S(g)={k\over 4\pi}\,\int
_{\Sigma}\,d^2z\,
Tr(\,g^{-1}\star \partial g\star g^{-1}\star  \bar \partial
g\,)-{k\over 12\pi}\,\int _{M}\,
Tr(\,g^{-1}\star dg\,)_\star ^3,
\ee
where $M$ is a three-dimensional manifold whose boundary is $\Sigma$,
and $g$ is a map from $\Sigma$ (or from its extension $M$) to the group $G$.
We assume that the coordinates $(z,\bar z)$ on the worldsheet $\Sigma$ are
noncommutative
but the extended coordinate $t$ on the manifold $M$  commutes with others:
\be
[z,\bar z]=\theta ,\quad [t,z]=[t,\bar z]=0.
\ee
We define the group-valued field $g$ by,
\be
g=e_\star ^{i\pi ^aT_a}= 1+i\pi ^aT_a+{1\over 2}(i\pi ^aT_a)_\star ^2+\cdots,
\ee
where the $T_a$'s are the generators of group $G$.

Inserting the $\star $-product of two group elements in the eq.
(\ref{action1}), we find the noncommutative Polyakov-Wiegmann identity,
\be\label{PW}
S(g\star h)= S(g) + S(h) + {1\over{16\pi}} \int d^2z Tr(g^{-1}\star
\bar\partial g \star
\partial h \star h^{-1}),
\ee
which is the same as ordinary commutative identity with
products replaced by $\star $-products.

Using the Polyakov-Wiegmann identity, we can show that the
action (\ref{action1}) is invariant under the following transformations:
\be\label{symm}
g(z,\bar z)\rightarrow h(z)\star  g(z,\bar z) \star  \bar h(\bar z).
\ee
The corresponding conserved currents
are, \bea
J(z) &=& {k\over{2\pi}}\bar\partial g\star  g^{-1},\nonumber\\
\bar J(\bar z) &=& {k\over{2\pi}}g^{-1} \star  \partial g.
\eea
By use of the equations of motion, we can show that these currents are
indeed conserved,
\be \l{EQMO}
\bar\p J(z)= \p \bar J(\b z)= 0.
\ee
We quantized this model by using Poisson brackets. To illustrate this method,
consider the following arbitrary action which is first order in time
derivative,
\be
S=\int dt A_i(\phi )\star {d\phi ^i\o {dt}},
\ee
Under the infinitesimal change of
fields $\phi ^i\rightarrow \phi ^i+\delta \phi ^i$,
one can calculate the variation of the action \cite{WITTEN},
\be\label{DELTAS}
\delta S=\int dt F_{ij}\star \delta \phi ^i\star {d\phi ^j\o {dt}},
\ee
in which $F_{ij}=\p _i A_j-\p _j A_i$.
The Poisson bracket of any two dynamical variables $X$ and $Y$ in the phase
space can be written as,
\be\label{pb}
[X,Y]_{PB}=\sum _{i,j} F^{ij}{\p X \o {\p \phi ^i}}{\p Y\o {\p \phi ^j}},
\ee
where $F^{ij}$ is the inverse of $F_{ij}$. The advantage of the above method is
that no explicit introduction of coordinates and momentum is necessary. Now,
consider the case for
the action (\ref{action1}), whose variation under $g \rightarrow g+\delta g$
leads to,
\be
\delta S={k\over{2\pi}}\int d\sigma\d\tau Tr(g^{-1}\star \delta g \star
{d \over
{d\sigma}}
(g^{-1}\star {dg\over{d\tau}})).
\ee

Let $(g^{-1}\star \delta g)=(g^{-1}\star \delta g)^a T_a$ and
$(g^{-1}\star {dg\over{d\tau}})=(g^{-1}\star {dg\over{d\tau}})^b T_b$,
where $T_a$ and $T_b$ are the group generators. So one can introduce
 $(g^{-1}\star \delta g)^a$ and $(g^{-1}\star {dg\over{d\tau}})^b$ as $\delta
\phi^i$ and ${d\phi^j \over{d\tau}}$ in (\ref{DELTAS}), respectively.
Therefore the $F_{ab}$ can be read as,
\be
F_{ab}=\delta_{ab} {k\over{2\pi}} {d \over {d\sigma}}
\ee
and its inverse is,
\be
F^{ab}=\delta^{ab} {{2\pi}\over k} ({d \over {d\sigma}})^{-1}
\ee

Now in (\ref{pb}) we take $X\equiv {{2\pi}\over k}J_a=Tr(T_a{dg \over
{d\sigma}}\star  g^{-1})$ and $Y\equiv {{2\pi}\over k} J_b=Tr(T_b{dg\over
{d\sigma '}}\star  g^{-1})$. To find the ${\p X \o {\p \phi ^i}}{\p Y\o {\p
\phi ^j}}$ term, we can evaluate the variation of $X$ and $Y$,
\be
\delta X \delta Y = \delta \phi^i \delta \phi^j {\p X \o {\p \phi ^i}}{\p
Y\o {\p \phi ^j}},
\ee
then drop the $\delta \phi^i \delta \phi^j $. After some algebra one
can find,
\be
[X,Y]={{2\pi}\over k}i\delta (\sigma -\sigma ')Tr[T_a,T_b]{dg\over
{d\sigma}}\star g^{-1}+{{2\pi}\over k}i\delta' (\sigma -\sigma ')Tr(T_aT_b),
\ee
or,
\be
[J_a(\sigma),J_b(\sigma')]=\delta (\sigma -\sigma ')if_{ab}^c J_c(\sigma)
+
{k\over {2\pi}}i\delta' (\sigma -\sigma ')\delta_{ab},
\ee
and a similar relation holds for commutation of $\b J$'s.

Note that in the above commutation relation, $\theta$ does not appear,
and it is just as commutative ordinary affine algebra with the same
central charge. The absence of $\theta$ has been expected, since the
currents are holomorphic by equations of motion and hence commutative in
the sense of $\star $-product.

Constructing the energy momentum tensor is also straightforward,
\be\label{TENSOR}
T(z)= {1 \over {k+N}} :J_i(z)J_i(z):+{1\over k}:J_0(z)J_0(z):,
\ee
where $J_i$'s are $SU(N)$ currents and $J_0$ is the $U(1)$ current
corresponding to the subgroups of $U(N)=U(1)\times SU(N)$.
Again the  products in (\ref{TENSOR}) are commutative because of
holomorphicity
of the currents. So the Virasoro algebra is also the same as usual
standard form and its central charge is unchanged,
\be
c={{kN^2+N}\over{k+N}}.
\ee
\Section{Gauged Noncommutative WZW Models}

In this section, we want to gauge the chiral symmetry (\ref{symm}) as,
\be \label{gtrans}
g(z,\bar z)\rightarrow h_L (z,\bar z)\star  g(z,\bar z) \star  h_R(z,\bar
z).
\ee
where $h_L$ and $h_R$ belong to $H$ some subgroup of $G$.
For finding the invariant action under the above transformation we need to add
gauge fields terms to the action (\ref{action1}) as follows,
\be
S(g,A,\b A)=S(g) + S_A +S_{\bar A} + S_2 + S_4,
\ee
where, $S(g)$ is the action (\ref{action1}) and,
\bea\label{gaugeterm}
S_A &=& {k\over 4\pi}\,\int
\,d^2z\,Tr(\,A_L \star \bar\partial g\star g^{-1}\,) \nn\\
S_{\bar A} &=& {k\over 4\pi}\,\int
\,d^2z\,Tr(\,\bar A_R\star g^{-1} \star \partial g\,) \nn\\
S_2 &=& {k\over 4\pi}\,\int
\,d^2z\,Tr(\,\bar A_R\star A_L\,) \nn\\
 S_4 &=& {k\over 4\pi}\,\int
\,d^2z\,Tr(\,\bar A_R\star g^{-1} \star  A_L \star g\,).
\eea
Gauge transformations for the gauge fields are,
\bea\label{atrans}
A_L &\rightarrow& h_L \star  (A_L + d) \star  h_L^{-1},\nn\\
A_R &\rightarrow& h_R^{-1} \star  (A_R + d) \star  h_R.
\eea

Using the Polyakov-Wiegmann identity (\ref{PW}), one can find,
\bea \l{AHAH}
S(g')&=&S(h_L\star g\star h_R)=S(h_L)+S(g)+S(h_R)+{k \o {4\pi}} \int d^2 z Tr \{
g^{-1}
h_L^{-1} \d h_Lg\db h_Rh_R^{-1}\cr &+&g^{-1}\d g \db h_R h_R^{-1}
+h_L^{-1}\d
h_L\db g g^{-1} \}_\star ,
\eea
where $\{\,\}_\star $ means all products inside brackets are $\star $-products.
The gauge fields terms (\ref{gaugeterm}), under
(\ref{gtrans}) and (\ref{atrans}) transform as,
\bea \l{SA}
S_A \rightarrow S'_A&=&S_A+{k \o {4\pi}} \int d^2z Tr \{A_L(h_L^{-1}\db
h_L+g\db h_Rh_R^{-1}g^{-1})\cr
& &-\d h_Lh_L^{-1}\db h_Lh_L^{-1}-h_L^{-1}\d
h_L\db g g^{-1}-h_L^{-1}\d h_Lg\db h_Rh_R^{-1}g^{-1} \}_\star
\nn\\
\l{SAB}
S_{\b A} \rightarrow S'_{\b A}&=&S_{\b A}+{k \o {4\pi}} \int d^2z Tr \{
\b A_R(\d h_R h_R^{-1}+
g^{-1}h_L^{-1}\d h_Lg)\cr
& & + \db h_Rh_R^{-1}g^{-1}h_L^{-1}\d h_Lg+\db h_Rh_R^{-1}g^{-1}\d
g+h_R^{-1}\db h_Rh_R^{-1}\d h_R \}_\star
\nn\\
S_2 \rightarrow S'_2&=&{k \o {4\pi}}\int d^2z Tr\{ h_R^{-1}\b
A_Rh_Rh_LA_Lh_L^{-1}+h_R^{-1}\b A_Rh_Rh_L\d h_L^{-1}\cr
& & +h_R^{-1}\db
h_Rh_LA_Lh_L^{-1}+h_R^{-1}\db h_Rh_L\d h_L^{-1} \}_\star \cr
S_4 \rightarrow S'_4&=&S_4+{k \o {4\pi}}\int d^2z Tr\{-h_R^{-1}g^{-1}
h_L^{-1}\d h_Lg\db h_R\cr
& & -g^{-1}h_L^{-1}\d h_Lg\b A_R+h_R^{-1}g^{-1}A_Lg\db h_R \}_\star .
\eea

To find an invariant action $S(g,A,\b A)$, we have to choose
constraints on the subgroup elements $h_L$ and $h_R$. The first consistent
choice is,
\be
h_R= h_L^{-1} \equiv h,
\ee
and yields to following transformations,
\bea \l{GFV}
g &\rightarrow& g'=h^{-1}\star g\star h \cr
A&\rightarrow&A'=h^{-1}\star (A\star h+\d h )\cr
\b A&\rightarrow&\b A'=h^{-1}\star (\b A \star  h+\db h).
\eea
The corresponding invariant action, called vector gauged WZW action,
is,
\bea \l{SVECTOR}
S_V(g,A,\b A)&=&S(g)+S_A-S_{\b A}+S_2-S_4\cr
&=&S(g)+{k \o {2\pi}} \int d^2z Tr\{A\db gg^{-1}-\b Ag^{-1}\d g+A\b
A-g^{-1}Ag\b A\}_\star .
\eea

The second choice is to take $h_L=h_R \equiv h$ with $h$ belonging to
an Abelian subgroup of $G$. In this case we find the following gauge
transformations,
\bea \l{GFA}
g &\rightarrow& g'=h\star g\star h \cr
A&\rightarrow&A'=h\star (A\star h^{-1}+\d h^{-1})\cr
\b A&\rightarrow&\b A'=h^{-1}\star (\b A \star  h -\db h),
\eea
with the so called axial gauged WZW action,
\bea \l{SAXIAL}
S_A(g,A,\b A)&=&S(g)+S_A+S_{\b A}+S_2+S_4\cr
&=&S(g)+{k \o {2\pi}} \int d^2z Tr\{A\db gg^{-1}+\b Ag^{-1}\d g+A\b
A+g^{-1}Ag\b A\}_\star .
\eea
There are some other choices for the gauged transformations, which can be
constructed obviously as the commutative case \cite{EXCEP}, however they are
less common and we will not discuss them here.

By integrating out the $A$ and $\b A$ from the actions (\ref{SVECTOR}) and
(\ref{SAXIAL}), in principle, we find the effective actions as
noncommutative non-linear $\sigma$-models.
We take here the noncommutative gauged axial $U(2)$ WZW action
(\ref{SAXIAL}), gauged by the subgroup $U(1)$ diagonally embedded in
$U(2)$. The group element of $U(2)$ is,
\be
g=\pmatrix{a_1&a_2\cr a_3&a_4},
\ee
with the following constraints,
\bea    \label{CONST}
a_1\star a_1^\dg+a_2\star a_2^\dg&=&1,\cr
a_3\star a_3^\dg+a_4\star a_4^\dg&=&1,\cr
a_1\star a_3^\dg+a_2\star a_4^\dg&=&0.
\eea
The gauge parts of the action (\ref{SAXIAL}) is,
\be     \label{SGAUGE}
S_{gauge}={k \o {2\pi}} \int d^2z \{A\sum _i\b\p a_i\star a_i^\dg+
\sum _ia_i^\dg\star \p a_i \b A+2A\b A+\sum_i A\star a_i\star \b A\star
a_i^\dg \}, \ee
To illustrate the integrating over the gauge fields $A$ and $\b A$, we
consider abbreviated notations as
follows,
\bea\label{INTEG}
\int {\cal D}A{\cal D}\b A e^{-S_{gauge}}&=&
\int {\cal D}A{\cal D}\b A e^{-\int d^2z(A\star {\cal O}\star \b A+b\star
\b A+A\star \b b)}\nn\\
&=&\int {\cal D}A{\cal D}\b A e^{-\int d^2z\big( (A+b')\star {\cal O}\star (\b
A+\b b')-b'\star {\cal O}\star \b b'\big)}, \eea
where
\bea \label{BB}
b'\star {\cal O}&=&b,\\
\label{BBB}
{\cal O}\star \b b'&=&\b b.
\eea
The result of integration would be,
\be       \label{EFFECT}
e^{-S_{eff}}=(\det {\cal O})^{-1/2}e^{\int d^2zb'\star \b b}e^{-S(g)}.
\ee
By comparing eq. (\ref{INTEG}) with (\ref{SGAUGE}), $b$ and $\b b$ could be read
as follows,
\bea
b&=&{{k}\over {2\pi}}\sum _i a_i^\dagger\star \p a_i\nn\\
\b b&=&{{k}\over {2\pi}}\sum _i \b \p a_i\star a_i^\dagger ,
\eea
and ${\cal O}$ can be read from quadratic terms of gauge fields $A$ and $\b A$
in (\ref{SGAUGE}). In fact
by using the Fourier transformation of Moyal $\star $-products of functions,
\be \label{FOUR}
f_1(x)\star f_2(x)\star \cdots \star f_n(x)=\int f_1(p_1)f_2(p_2)\cdots f_n(p_n)
e^{i\sum _{i<j} p_i\wedge p_j}e^{i\sum _i p_i x}dp_1dp_2\cdots
dp_n,
\ee
the explicit Fourier transform of ${\cal O}$ is as follows,
\bea
{\cal O}(p_1,p_2)=2 e^{i(p_1\wedge p_2)} \delta (p_1+p_2)
&+&\int a_i(p_3)a_i^\dg (p_4) e^{i(p_1\wedge p_2+p_1\wedge p_3+p_1\wedge p_4
+p_2\wedge p_4+p_3\wedge p_4+p_3\wedge p_2)}\nn\\
& &\times \delta (p_1+p_2+p_3+p_4)
dp_3dp_4.
\eea
To find $b'$, we need to inverse the ${\cal O}$ operator in (\ref{BB}), and
this is equivalent to solve Fourier transform of (\ref{BB}) which is an
integral equation as follows,
\be \label{INTEQ}
2b'(p)+\int b'(p-p_1-p_2)a_i(p_1)a_i^\dg (p_2)
e^{-i(p\wedge p_2+p_1\wedge p+p_2\wedge p_1)}
dp_1dp_2=b(p),
\ee
where
\be
b(p)=i \int a_i^\dg (p-p_1)a_i(p_1)p_1
e^{i p\wedge p_1}dp_1.
\ee
To solve eq. (\ref{INTEQ}), we expand the $b'$ and exponential factors in
terms of $\theta$,
\be \label{ITER}
b'(p)=b'_0(p)+\theta b'_1(p)+\theta ^2 b'_2(p)+\cdots .
\ee
In zero-th order of $\theta$, one finds,
\be \l{BSEFR}
b'_0(z, \b z)={1 \over 4}a_i ^\dagger \p a_i,
\ee
and in first order of $\theta$,
\be \l{BAWAL}
b'_1(z, \b z)={1 \over {4}}(
+4\p \p a_i \b \p a_i^\dg
-4\b \p \p a_i \p a_i^\dg).
\ee

In obtaining the above expressions, we have used the unitarity
conditions
(\ref{CONST}) and the equations of motion (\ref{EQMO}).
The effective action arising from the $\int d^2z b'\star \b b$ term in
eq. (\ref{EFFECT}) could be found as a power series in
$\theta$,
\be
S_{eff}=S(g)+{1 \o 2}Tr \ln ({\cal O}) +S_{eff}^{(0)}+\theta
S_{eff}^{(1)}+\cdots , \ee
in which
\footnote{The term ${1\o 2}Tr \ln ({\cal O})$ gives the
noncommutative effective action for dilaton field.}
\bea \l{ACTIONS}
S_{eff}^{(0)}&=&-{{k}\over {8\pi}}\int d^2z a_i^\dagger \p a_i \b \p
a_j a_j^\dagger\nn\\
S_{eff}^{(1)}&=&-{{k}\over {8\pi}}\int d^2z \big( (
\p a_i^\dagger \p  a_i+ a_i^\dg \p \p a_i)
(\b\p\b \p a_j  a_j^\dg+\b \p a_j \b \p a_j^\dg)
+ a_i^\dg\p a_i(\p \b\p a_j\b \p a_j^\dg-\b \p\b\p a_j\p a_j^\dg) \nn \\
& &-
(\b \p \p a_i \p a_i^\dg
-\p \p a_i \b \p a_i^\dg ) \b \p a_ka_k^\dg \big) .
\eea
By looking at equations (\ref{BSEFR}), (\ref{BAWAL}) and (\ref{ACTIONS}), we
suggest the following exact forms for $b'(z,\b z)$ and $S_{eff}$,
\bea \l{EXACT}
b'(z, \b z)&=&{1 \over 4}\p a_i \star a_i ^\dagger \nn\\
S_{eff}&=&S(g)+{1 \o 2}Tr \ln ({\cal O})
-{{k}\over {8\pi}}\int d^2z \p a_i \star  a_i^\dagger \star \b \p
a_j \star a_j^\dagger .
\eea
We have to fix the gauge freedom by a gauge fixing condition
on $a_i$'s. Under infinitesimal axial gauge transformations (\ref{GFA}),
we find,
\be
a_i'=a_i+a_i\star \epsilon+\epsilon \star a_i ,
\ee
in which $\epsilon$ is the infinitesimal parameter of the gauge transformation.
We can find ( at least perturbatively in $\theta$ ) some $\epsilon$ such that
${\Re (a_1')}=0$ and one may take this relation as the gauge fixing
condition. It is worth mentioned that the $U(2)/U(1)$ model after applying
all conditions gives us a three dimensional non-linear
noncommutativre $\sigma$-model.

\Section{Conclusions}

We found current algebra and energy-momentum tensor of noncommutative WZW
models are the same as corresponding quantities in commutative WZW models.
By gauging the chiral symmetry of noncommutative WZW model, we showed that
different gauge actions could be constructed. After integrating out the
gauge fields from the axial gauged noncommutative WZW model
, we observed that the final result is a perturbative noncommutative
non-linear $\sigma$-model. In general, integrating over the quadratic action
demands inversing the quadratic operator ${\cal O}$, but the inversion
of operators in noncommutative space
is not obvious and it is natural to find an infinite series ( in powers of
$\theta$ ) for the
$\sigma$-model action, however, we suggested an exact form for the effective
action containing $\star$-products.
Applying the gauge condition will reduce this effective action to a
$\sigma$-model on a three dimensional noncommutative target space.
The geometrical study of this target space which may contains singular
structure will be interesting.
\vskip .2 in

\noindent{\bf{Acknowledgement}}
\vskip .1 in
We would like to thank F. Ardalan and V. Karimipour for useful discussions.

\end{document}